\documentclass[10pt,twocolumn,letterpaper]{article}

\usepackage{cvpr}
\usepackage{times}
\usepackage{epsfig}
\usepackage{graphicx}
\usepackage{subcaption}
\usepackage{amsmath}
\usepackage{amssymb}
\usepackage{booktabs}
\usepackage{tabularx}
\usepackage{makecell}
\usepackage{multirow}
\usepackage{enumitem}
\usepackage[table]{xcolor}
\definecolor{lightgray}{gray}{0.92}
\usepackage{caption}
\makeatletter
\renewcommand\subsection{\@startsection{subsection}{2}{\z@}%
  {0.8\baselineskip}
  {0.3\baselineskip}
  {\normalfont\large\bfseries}}
\makeatother

\usepackage[pagebackref=true,breaklinks=true,letterpaper=true,colorlinks,bookmarks=false]{hyperref}

\cvprfinalcopy 

\def\cvprPaperID{****} 

\ifcvprfinal\fi
\begin{document}

\title{MMAudio-LABEL: Audio Event Labeling via Audio Generation for Silent Video}  
\author{
Kazuya Tateishi$^{1}$,
Akira Takahashi$^{1}$,
Atsuo Hiroe$^{1}$,
Hirofumi Takeda$^{1}$,\\
Shusuke Takahashi$^{1}$,
Yuki Mitsufuji$^{1,2}$\\
$^{1}$Sony Group Corporation, Japan \quad
$^{2}$Sony AI, USA
}

\maketitle
\thispagestyle{empty}

\abstract{}
Recent advances in multimodal generation have enabled high-quality audio generation from silent videos. Practical applications, such as sound production, demand not only the generated audio but also explicit sound event labels detailing the type and timing of sounds. One straightforward approach involves applying a standard sound event detection to the generated audio. However, this post-hoc pipeline is inherently limited, as it is prone to error accumulation. To address this limitation, we propose MMAudio-LABEL (\textbf{LA}tent-\textbf{B}ased \textbf{E}vent \textbf{L}abeling), an event-aware audio generation framework built on a foundational audio generation model as its backbone that jointly generates audio and frame-aligned sound event predictions from silent videos. We evaluate our method on the Greatest Hits dataset for onset detection and 17-class material classification. Our approach improves onset-detection accuracy from 46.7\% to 75.0\% and material-classification accuracy from 40.6\% to 61.0\% over baselines. These results suggest that jointly learning audio generation and event prediction enables a more interpretable and practical video-to-audio synthesis.

\normalfont

\section{Introduction}
Generating realistic and accurate sounds from silent videos is a long-standing goal with applications in content creation, immersive media, and human-computer interaction.
This task is challenging because the model must infer semantically plausible acoustic events from visual cues and align them precisely over time.
Recent progress in multimodal generative modeling has led to substantial advances in video-to-audio (V2A) synthesis~\cite{cheng2025mmaudiotamingmultimodaljoint,liu2025tellhearvideo, ren2025stav2avideotoaudiogenerationsemantic, zhang2024foleycrafterbringsilentvideos}.
Among these approaches, MMAudio~\cite{cheng2025mmaudiotamingmultimodaljoint} demonstrates strong audio-visual synchronization by extracting semantic cues and fine-grained temporal dynamics from video.

Most existing V2A models primarily focus on improving generation quality and conditioning-based controllability~\cite{cheng2025mmaudiotamingmultimodaljoint,du2023conditionalgenerationaudiovideo,fang2026acfoley, jeong2024readwatchscreamsound,liu2025tellhearvideo, zhang2024foleycrafterbringsilentvideos}. However, they do not explicitly expose event-level information aligned with the video timeline. In practical content production, creators often require explicit guidance on \emph{what} sound events occur and \emph{when} to efficiently place and verify sounds in silent videos.

We address a twofold task: generating high-quality audio from a silent video and identifying the types and timings of the corresponding sound events.
Since existing V2A models lack built-in event detection capabilities, one straightforward solution involves attaching a conventional sound event detection model~\cite{kong2020pannslargescalepretrainedaudio} to the generated audio.
Although such post-hoc pipelines can detect event types and timing from audio, they are decoupled from the generation process, discard visual context, and may fail to distinguish acoustically similar events (e.g., footsteps vs.\ knocking).
Another line of work~\cite{10889172} uses object detection to extract discrete visual cues, and then generates audio conditioned on these cues.
Although visual information can be utilized, event classes defined for image recognition tasks do not always correspond to sound events.

\begin{figure}[tb]
  \centering
  \includegraphics[width=\linewidth]{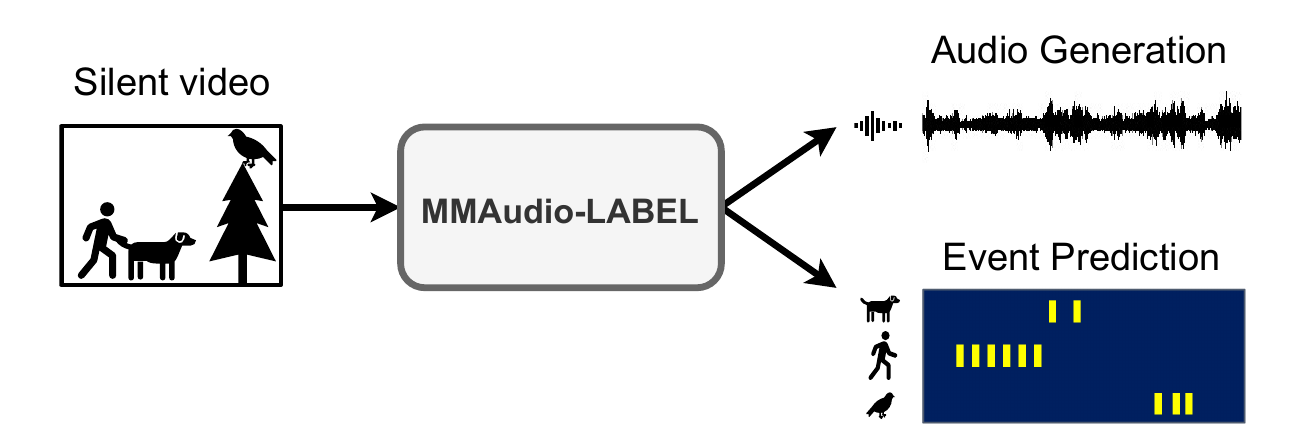}
  \caption{\small Overview of MMAudio-LABEL.}
  \label{fig:concept}
\end{figure}

To bridge generation and sound event interpretability, we propose \textbf{MMAudio-LABEL}, an event-aware V2A framework that outputs both a generated audio and time-resolved sound event labels (Figure~\ref{fig:concept}).
The proposed framework enables semantically and temporally aligned frame-level event labeling.
Building on the success of MMAudio~\cite{cheng2025mmaudiotamingmultimodaljoint} as a foundation model, which has been demonstrated in downstream tasks like sound separation with MMAudioSep~\cite{takahashi2025mmaudioseptamingvideotoaudiogenerative}, we explore its capabilities in event labeling.
We evaluate our approach on the Greatest Hits dataset~\cite{owens2015visually}, where percussive interactions provide clear onsets and material labels to precisely assess event timing and material class.

\begin{figure*}[!t]
  \centering
  \captionsetup{width=0.9\textwidth,font=small} 

  \begin{subfigure}[t]{0.9\textwidth}
    \centering
    \includegraphics[width=\linewidth,clip,trim=0 140mm 0 0]{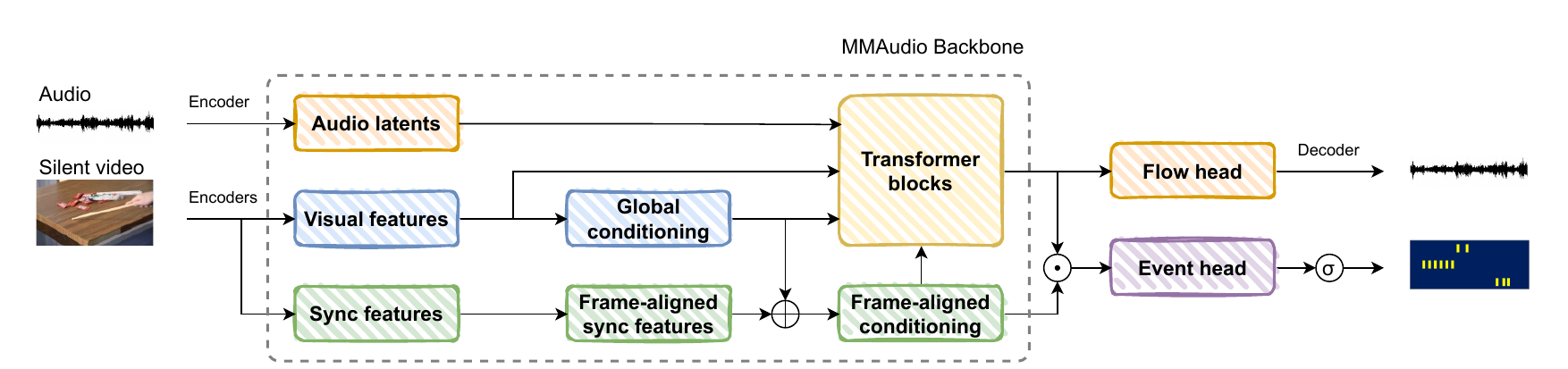}
    \caption{Parallel Heads (Predictive approach): parallel flow and event heads.}
    \label{fig:parallel_heads}
  \end{subfigure}

  \vspace{1mm}

  \begin{subfigure}[t]{0.9\textwidth}
    \centering
    \includegraphics[width=\linewidth,clip,trim=0 35mm 0 95mm]{fig/Structures.drawio_Shrink.pdf}
    \caption{Joint Heads (Generative approach): unified audio-event heads with split output.}
    \label{fig:joint_heads}
  \end{subfigure}

  \caption{Proposed architectures. Both models build on MMAudio’s flow-prediction network, where video conditions and target latents interact through transformer blocks. (a) The target latents are audio latents, and separate heads predict audio and events.
  (b) The target latents are a concatenation of audio and event latents, and a single head jointly predicts them.}
  \label{fig:architectures}
\end{figure*}

The contributions of this study are as follows:
\begin{itemize}[leftmargin=1em,topsep=1pt,itemsep=0pt,parsep=0pt,partopsep=0pt]
  \item We propose \textbf{MMAudio-LABEL}, an event-aware V2A framework that jointly models audio generation and sound event labeling in a shared latent space.
  \item We explore two distinct architectures and demonstrate the superiority of a unified generative approach.
  \item We demonstrate our method as an effective downstream task for a foundation model, achieving significant gains through finetuning.
\end{itemize}

\section{MMAudio-LABEL}

\subsection{Overview}
Given a silent video, our goal is to output semantically and temporally aligned audio and event information. MMAudio-LABEL builds on MMAudio’s strong audio-visual synchronization capability and extends it with an explicit event prediction mechanism. We jointly learn audio generation and frame-level event prediction, rather than relying on a post-hoc classifier, which encourages temporally structured representations that support both objectives.

\subsection{Model Architecture}
Figure~\ref{fig:architectures} illustrates our main architectures, which follow the overall structure of MMAudio~\cite{cheng2025mmaudiotamingmultimodaljoint}.
Given an input video, we encode semantic content with a visual encoder~\cite{radford2021learningtransferablevisualmodels} and perform audio generation in the audio-latent space using a multimodal transformer.
To achieve high temporal resolution, we also extract high-frame-rate synchronization features using Synchformer~\cite{iashin2024synchformerefficientsynchronizationsparse}, and condition each transformer layer on them.
This enables temporally fine-grained generation aligned with the video dynamics.

We aim to output frame-level audio latents and a multi-class event logit map at the same temporal resolution.
To achieve this, we explore two architectures: (a) \emph{Parallel Heads} and (b) \emph{Joint Heads}.

\textbf{(a) Parallel Heads: parallel flow and event heads.}
As a straightforward approach, we leverage MMAudio’s semantic understanding and synchronization features by introducing a parallel-head design (Figure~\ref{fig:parallel_heads}): one head for audio generation and another for event prediction.
To couple event prediction more tightly with the generated audio representation, we concatenate the output of the final audio-generation transformer block with the event-head input.

\textbf{(b) Joint Heads: unified audio-event heads with split output.}
As a unified approach, we incorporate event prediction into the generative process via a joint-head design (Figure~\ref{fig:joint_heads}): a single head jointly predicts audio latents and event logits within the same generation trajectory, preserving temporal resolution and class structure.

During training, we treat event logits as continuous variables and add timestep-dependent noise to obtain event latents.
We then concatenate the audio latents with the noised event latents and feed the joint latents into the multimodal transformer.
Finally, we split the output of the flow head into two parts corresponding to the dimensionalities of the audio latents and the event logit maps.

During testing, we start with noisy audio latents and noisy event logits and iteratively denoise them along the generation trajectory.
The audio-latent component is decoded by a variational autoencoder~\cite{kingma2022autoencodingvariationalbayes} into spectrograms, which are then fed into a vocoder~\cite{lee2023bigvganuniversalneuralvocoder} to synthesize the final waveform, while the event-logit component is passed through a sigmoid function to obtain frame-level per-class probability estimates.

\subsection{Objective Functions}
We train the flow head using a conditional flow matching objective~\cite{lipman2023flowmatchinggenerativemodeling,tong2024improvinggeneralizingflowbasedgenerative} for generative modeling. We optimize the model parameters $\theta$ of the time-dependent velocity field $v_{\theta}$ by minimizing the following losses:
{
\setlength{\abovedisplayskip}{6pt}
\setlength{\belowdisplayskip}{6pt}
\setlength{\abovedisplayshortskip}{6pt}
\setlength{\belowdisplayshortskip}{6pt}
\begin{equation}
\mathcal{L}_{\text{flow}} =
\mathbb{E}_{t,\,q(x_0),\,q(x_1,\mathbf{C})}
\left\| v_{\theta}(t,\mathbf{C},x_t) - (x_1 - x_0) \right\|^2,
\label{eq:lflow}
\end{equation}
}
where $t \in [0, 1]$ is the continuous timestep, $q(x_0)$ is the standard normal distribution, $\mathbf{C}$ is video condition, and $q(x_1,\mathbf{C})$ is sampled from the training data. We construct the interpolated sample as $x_t = t x_1 + (1-t) x_0$, that is, a linear interpolation between the noise sample $x_0$ and the target latent $x_1$.

For Parallel Heads, we apply a Binary Cross-Entropy (BCE) loss $\mathcal{L}_{\text{bce}}$ to the event-class output and optimize the following two objectives:
{
\setlength{\abovedisplayskip}{6pt}
\setlength{\belowdisplayskip}{6pt}
\setlength{\abovedisplayshortskip}{6pt}
\setlength{\belowdisplayshortskip}{6pt}
\begin{equation}
\mathcal{L} = \mathcal{L}_{\text{flow}} + w\,\mathcal{L}_{\text{bce}},
\label{eq:jointloss}
\end{equation}
}
where $w$ denotes the weighting factor.

For Joint Heads, when applying the logit transform, we add a small $\epsilon$ to the binary (0/1) ground-truth labels to avoid numerical issues (e.g., division by zero or $\log(0)$).

\section{Experiments}
\subsection{Dataset and Downstream Tasks}
We used the Greatest Hits dataset, consisting of videos in which a drumstick hits various materials. Each video includes the corresponding audio track and material labels indicating the hit surface.
We evaluated the proposed method on two tasks, onset detection and material classification, to assess its performance on predicting sound timing and types, respectively.

\textbf{Onset Detection}: Onset detection aims to identify the time instants (onsets) at which the drumstick hits an object from silent video. We followed the data split used in ~\cite{du2023conditionalgenerationaudiovideo}, training on the training set and evaluating on the same test set. We evaluated temporal alignment performance using count matching for the number of hits per test sample, accuracy (Acc), and average precision (AP), with an onset tolerance of $\pm 0.1$\,s. Additionally, we assessed the quality of the generated audio using Mel Cepstral Distortion (MCD). We considered two baselines: CondFoley~\cite{du2023conditionalgenerationaudiovideo} and MMAudio~\cite{cheng2025mmaudiotamingmultimodaljoint} (small-16k model).

\textbf{Material Classification}: Material classification aims to recognize the material being hit from silent video. We considered all 17 material classes in the Greatest Hits dataset. For training, we constructed the training set using only clips that contain a single material label. For testing, we built a test set from the official test split, consisting of 11 clips of 2\,s per class. For materials not present in the test split, we used samples from the validation split. As a baseline, we trained a VGGish Classifier on the Greatest Hits dataset following ~\cite{du2023conditionalgenerationaudiovideo}. This model served as an audio-based classifier. Because the VGGish Classifier performs clip-level classification, we adopted clip-level accuracy (Acc) as the evaluation metric. For our proposed models, the final clip-level prediction is determined by the most frequently predicted class within that clip.

\subsection{Implementation Details}
In our models, audio was decoded to 16\,kHz waveforms from 20-dimensional latents. For parallel heads, the event head consisted of a three-layer MLP.
For the joint heads, we augmented the audio latent dimensionality with the number of event classes. Accordingly, we used 21-dimensional latents for onset detection and 37-dimensional latents for material classification, and split them into audio and event components after the flow head.
The model was trained on an NVIDIA RTX A6000 GPU using the AdamW optimizer with an initial learning rate of $1\times10^{-4}$. We applied a linear warm-up for the first 1{,}000 steps, followed by a scheduled learning-rate decay starting at 50{,}000 steps to $1\times10^{-5}$. Training was performed for 100{,}000 iterations, with a batch size of 16.
We set the objective weighting factor and the small constant to prevent division by zero to $w=1$, $\epsilon=1\times10^{-5}$ respectively.

Since our method was trained on 8-second clips, we avoided a mismatch in the input duration at test time. For the 2-second test set, we looped and concatenated each clip to 8\,s before running inference. We used the predictions corresponding to the first 2\,s for evaluation.

\subsection{Results}
For onset detection, both the parallel and joint heads substantially outperformed prior methods (Table~\ref{tab:onset_results}). Among them, the joint heads is superior, improving upon the parallel heads in all temporal alignment metrics while also achieving a better MCD score for audio quality. The finetuned joint heads achieved the highest performance on all metrics.

\begin{figure}[tb]
  \centering
  \begin{subfigure}[t]{0.49\linewidth}
    \centering
    \includegraphics[width=\linewidth]{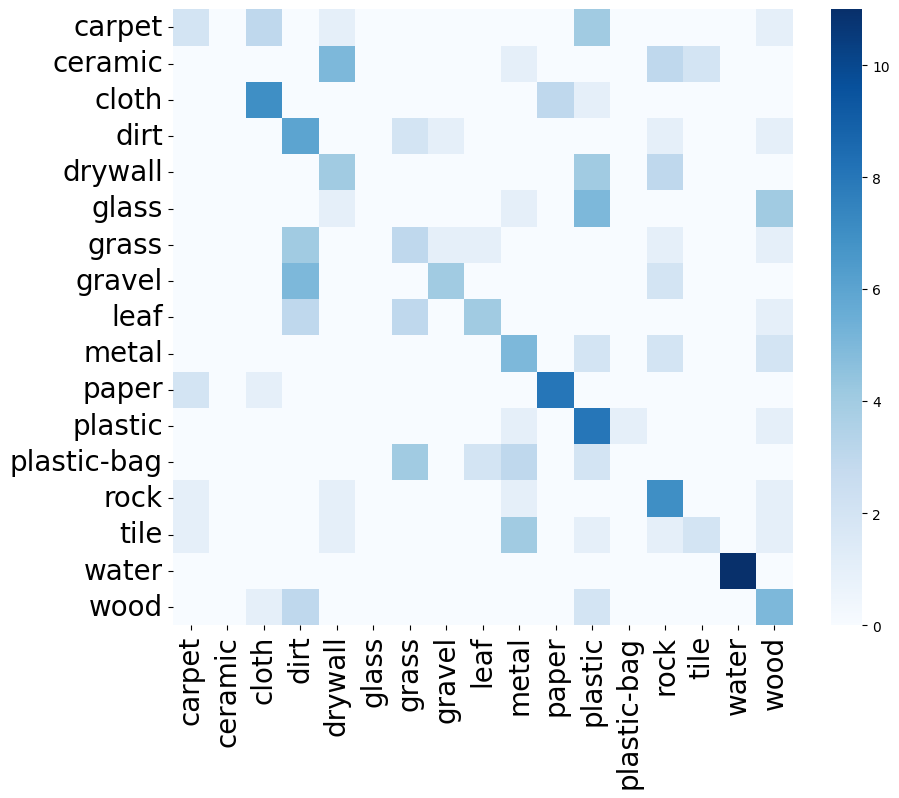}
    \caption{VGGish Classifier (Baseline)}
    \label{fig:conf-baseline}
  \end{subfigure}\hfill
  \begin{subfigure}[t]{0.49\linewidth}
    \centering
    \includegraphics[width=\linewidth]{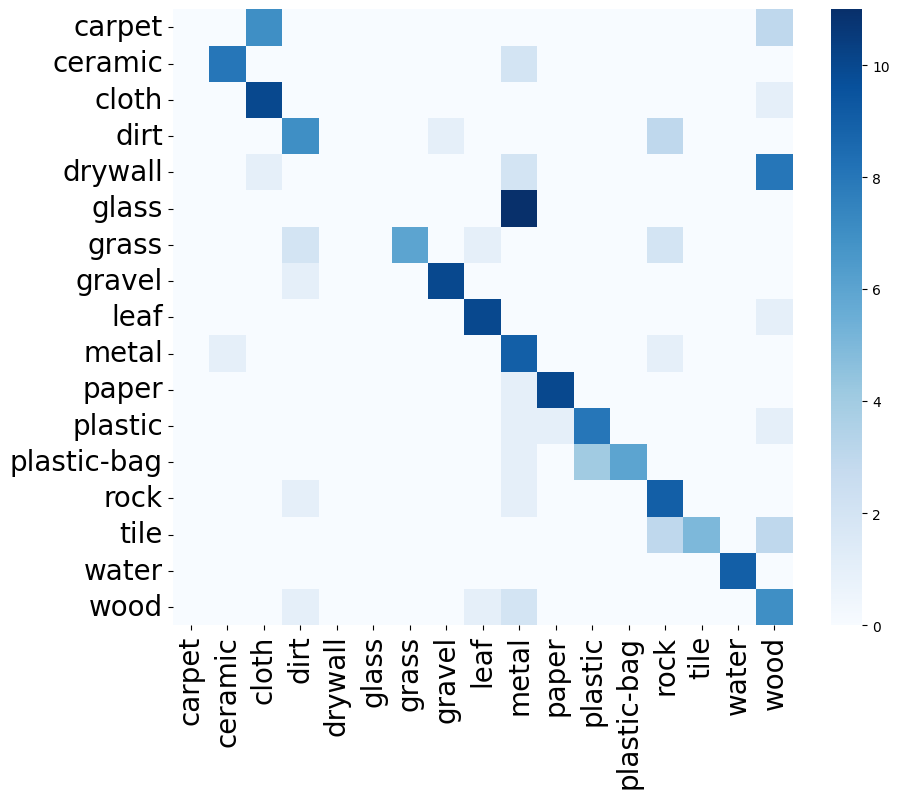}
    \caption{MMAudio-LABEL (Ours)}
    \label{fig:conf-ours}
  \end{subfigure}
  \caption{Confusion matrices for classification task}
  \label{fig:confusion}
\end{figure}

A similar trend was observed for material classification (Table~\ref{tab:material_results}). The finetuned joint heads achieved a significant gain, followed by the joint heads (scratch). This suggests that a joint latent representation that integrates audio generation and event prediction enables the model to not only extract more informative visual cues but also boost the quality of the audio generation itself. A comparison of the confusion matrices for the VGGish Classifier and our Joint Heads (finetune) reveals that our method achieves superior overall performance, even against the VGGish Classifier using ground-truth audio (Figure~\ref{fig:confusion}).
Although recognizing materials with less distinctive shapes, such as carpet, drywall, and glass remains a challenge, finetuning improves accuracy from 51.9\% to 61.0\%. This indicates that leveraging a pretrained model learned from large-scale multimodal data can further enhance specific downstream tasks.

\begin{table}[t]
\centering
\caption{Onset detection results.}
\label{tab:onset_results}
\resizebox{\columnwidth}{!}{%
\begin{tabular}{l l c c c c}
\toprule
Model & Training & \makecell[c]{Count\\match(\%) $\uparrow$} & Acc(\%) $\uparrow$ & AP(\%) $\uparrow$ & MCD $\downarrow$ \\
\midrule
CondFoley & Scratch & 30.0 & 46.7 & 63.5 & 8.85\\
MMAudio small-16k & Pretrain & 20.6 & 24.8 & 65.1 & 9.95\\
\midrule
\multirow{4}{*}{\makecell[l]{MMAudio-LABEL (Ours)\\\hspace{0.8em}Event Head Only\\\hspace{0.8em}Parallel Heads\\\hspace{0.8em}Joint Heads}}
 &  &  &  &  & \\
 & Scratch & 17.5 & 22.0 & 74.4 & no audio\\
 & Scratch & 49.0 & 70.5 & 89.3 & 8.31\\
 & Scratch & 53.1 & 71.3 & 90.0 & 8.27\\
\rowcolor{lightgray}
\hspace{0.8em}Joint Heads (finetune)$^{*}$ & Finetune & \textbf{54.6} & \textbf{75.0} & \textbf{91.6} & \textbf{8.22}\\
\bottomrule
\end{tabular}}
\vspace{1mm}
\noindent\hfill{\footnotesize $^{*}$finetuned from small-16k checkpoint.}
\end{table}

\begin{table}[t]
\centering
\caption{Material classification results.}
\label{tab:material_results}
\resizebox{\columnwidth}{!}{%
\begin{tabular}{l c c c c}
\toprule
Model & Training & Input & Output & \makecell[c]{Acc(\%) $\uparrow$ } \\
\midrule
VGGish Classifier & Scratch & audio & label & 40.6 \\
\midrule
\multirow{4}{*}{\makecell[l]{MMAudio-LABEL (Ours)\\\hspace{0.8em}Event Head Only\\\hspace{0.8em}Parallel Heads\\\hspace{0.8em}Joint Heads}}
 &  &  &  &  \\
 & Scratch & \makecell[l]{visual} & label & 39.0 \\
 & Scratch & \makecell[l]{visual} & \makecell[l]{label + audio} & 43.9 \\
 & Scratch & \makecell[l]{visual} & \makecell[l]{label + audio} & 51.9 \\
\rowcolor{lightgray}
\hspace{0.8em}Joint Heads (finetune)$^{*}$ & Finetune & \makecell[l]{visual} & \makecell[l]{label + audio} & \textbf{61.0} \\
\bottomrule
\end{tabular}}
\vspace{1mm}
\noindent\hfill{\footnotesize $^{*}$finetuned from small-16k checkpoint.}
\end{table}
\vspace{-2mm}

\section{Conclusion}
In this paper, we proposed MMAudio-LABEL, a framework for jointly modeling audio generation and event prediction from silent video. Experiments demonstrated consistent improvements over prior baselines on both onset detection and material classification. In particular, the joint-heads design achieved the best overall performance across both temporal alignment and audio quality metrics, and this performance was further enhanced through finetuning. This suggests that integrating event information with the prior knowledge into the audio latent space yields a more effective representation.
Overall, these results highlight the benefits and broader potential of joint modeling and its success as a downstream application of the foundation model.


{\small
\bibliographystyle{ieee}
\bibliography{main}
}

\end{document}